\documentclass[manuscript]{acmart}
\usepackage{tabularx}
\usepackage{float}
\usepackage{vtable}
\usepackage{multirow}
\usepackage{subcaption}

\AtBeginDocument{%
  \providecommand\BibTeX{{%
    \normalfont B\kern-0.5em{\scshape i\kern-0.25em b}\kern-0.8em\TeX}}}

\setcopyright{acmlicensed}
\copyrightyear{2018}
\acmYear{2018}
\acmDOI{XXXXXXX.XXXXXXX}

\acmConference[Conference acronym 'XX]{Make sure to enter the correct conference title from your rights confirmation email}{June 03--05, 2018}{Woodstock, NY}

\acmISBN{978-1-4503-XXXX-X/18/06}

\begin{document}

\title{EyeBrain: Left and Right Brain Lateralization Activity Classification Through \\ Pupil Diameter and Fixation Duration}

\author{Ko Watanabe}
\orcid{0000-0003-0252-1785}
\email{ko.watanabe@dfki.de}
\affiliation{%
  \institution{RPTU Kaiserslautern-Landau {\&} DFKI GmbH}
  \city{Kaiserslautern}
  \country{Germany}
}

\author{Pooja Pol}
\orcid{}
\email{}
\affiliation{%
  \institution{RPTU Kaiserslautern-Landau}
  \city{Kaiserslautern}
  \country{Germany}
}

\author{Nicolas Großmann}
\orcid{1234-5678-9012}
\email{trovato@corporation.com}
\affiliation{%
  \institution{RPTU Kaiserslautern-Landau {\&} DFKI GmbH}
  \city{Kaiserslautern}
  \country{Germany}
}

\author{Shoya Ishimaru}
\orcid{0000-0002-5374-1510}
\email{ishimaru@omu.ac.jp}
\affiliation{%
  \institution{Osaka Metropolitan University {\&} DFKI Lab Japan}
  \city{Osaka}
  \country{Japan}
}

\author{Andreas Dengel}
\orcid{0000-0002-6100-8255}
\email{andreas.dengel@dfki.de}
\affiliation{%
  \institution{RPTU Kaiserslautern-Landau {\&} DFKI GmbH}
  \city{Kaiserslautern}
  \country{Germany}
}

\renewcommand{\shortauthors}{Watanabe and Pol et al.}


\begin{abstract}
The relationship between brain lateralization and cognitive functions is well-documented. The left hemisphere primarily handles tasks such as language and arithmetic, while the right hemisphere is involved in creative activities like drawing and music perception. Eye-tracking technology has shown the potential to reveal cognitive states by measuring ocular metrics such as pupil diameter and fixation duration. However, the ability to distinguish lateralized brain activity using these ocular metrics remains underexplored. Here, we demonstrate that pupil diameter and fixation duration can effectively classify left and right brain hemisphere activities. We obtained a considerably high classification performance, with an F1 score of 0.894. The results suggest that ocular metrics are robust indicators of lateralized brain activity and can be applied in cognitive monitoring and neurorehabilitation. Our future work expands on this by integrating these methods into real-time applications \textit{EyeBrain}, potentially broadening their use across various cognitive and neurological domains.
\end{abstract}




\begin{teaserfigure}
  \centering
  \includegraphics[width=0.85\textwidth]{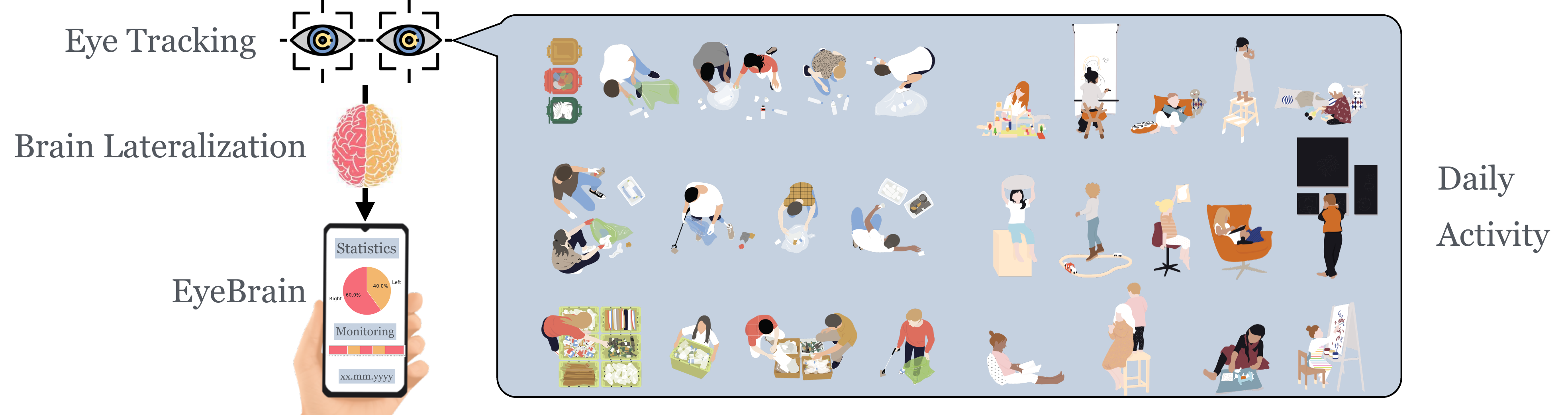}
  \caption{Concept design of \textit{EyeBrain} project. Eye-tracking technology makes statistical information on \textit{brain lateralization} accessible and easy to understand. Users can understand brain activities as an activity monitoring application.}
  \label{fig:teasure}
\end{teaserfigure}

\maketitle

\section{Introduction}
\label{sec:intro}
The brain function of humans is closely related to features observable through their eyes~\cite{london2013retina, oliva2018pupil, joshi2020pupil}.
Pupil diameter, for example, correlates with the activity of the locus coeruleus~\cite{murphy2014pupil, joshi2016relationships, nobukawa2021pupillometric}, which is a brain region critical for managing functions of both long-term and short-term memory~\cite{kahneman1966pupil, kucewicz2018pupil}.
Not only from eyes to detect brain activity, but also there is research on detecting left and right saccades direction from Electroencephalography (EEG) data achieving classification with 94.8\% using InceptionTime network~\cite{kastrati2021using}.
Thus, brain and eye activity have been found to correlate in various ways.
Due to these exciting observations, eye-tracking has been used to detect fatigue~\cite{abdulin2015user, tag2019continuous}, mental workload~\cite{kosch2018look, grimmer2021cognitive}, task difficulty~\cite{duchowski2018task, cho2021rethinking, bacchin2023gaze}, mind wandering~\cite{brishtel2020mind, hutt2021breaking}, lie detection~\cite{mann2002suspects, leal2008blinking, grootjen2024uncovering}, and other cognitive activity recognition.

Our interest aligns to estimate brain activity from the eyes, especially in the left and right hemispheres.
The concept of left and right asymmetry or so-called \textit{brain lateralization}~\cite{rogers2021brain} explains that brain hemispheres each have different roles~\cite{walker1980lateralization, toga2003mapping, neubauer2020evolution}.
The left hemisphere's role is to process language~\cite{toga2003mapping, neubauer2020evolution} and arithmetic~\cite{semenza2006math, funnell2007calculating, pinel2010beyond}.
On the other hand, the right hemisphere's role is to process drawing~\cite{kirk1989hemispheric, nikolaenko2003artistic} and listening to music~\cite{gordon1983music, peretz2005brain}.
Based on these findings, we focus on whether it is possible to estimate right-left brain hemisphere activation from eye movements by measuring the pupil during human right-left brain hemisphere-related behaviors.

Figure~\ref{fig:teasure} shows the application design goal of the \textit{EyeBrain} concept.
Similar to activity monitoring in smartphone applications, we make accessible \textit{brain lateralization} statics while using eye-tracking.
The application aims to support which lateralization activity, especially left or right brain activity, has been active in daily life.
The statistics will help users monitor the balance of their cognitive tasks in daily life and enable them to understand strengths and weaknesses from statistical review.
From the statistical review, users can choose to work on either logical and imaginative activities in balance or to focus on one another.
This segmentation of behavior at the brain region level allows individuals to visualize their brain lateralization statistics and motivates them to train for each brain region.

It monitors daily brain lateralization statistics and enables tracking of monthly or yearly activity, which may also be helpful for future career decisions.
The user can understand how often they focus on right or left brain activity and their likes and strengths.
Also, mentors like teachers can advise the activity to work on by looking at daily, weekly, or yearly monitoring.
The data is also simple enough, so it is expected to provide a more comprehensive picture of daily cognitive behavior.

In order to achieve this goal, the biggest challenge is to implement the generalized and explainable machine-learning model for recognizing \textit{brain lateralization} through eye-tracking.
To do so, we only use pupilometry information in the eye-tracking.
In our case, pupilometry corresponds to the features like pupil diameter and fixation.
We removed features like gaze coordinates or saccade direction from our input features since these features strongly correlate with the user interface and the task of activities.
Our primary goal is to measure \textit{brain lateralization} from features less affected by the user interface, a crucial aspect of our research.

In this study, we aim to recognize left and right \textit{brain lateralization} from the eye-tracking dataset.
As mentioned previously, the eyes and brain are strongly correlated.
In particular, previous studies have shown that brain activity is affected and altered by the pupil~\cite{london2013retina, oliva2018pupil, joshi2020pupil}.
Using these features, we challenge to estimate left and right \textit{brain lateralization} using eye-tracking.
In this study, we use eye-tracking features, mainly pupil diameter and fixation duration.
To summarize the objectives, our focus research questions are as follows:

\begin{itemize}
    \item[RQ1] Can we recognize right and left \textit{brain lateralization} from pupil diameter and fixation duration?
    \item[RQ2] What features effectively estimate right and left \textit{brain lateralization} activities from pupil diameter and fixation duration?
\end{itemize}

\section{Related Work}
In this section, we first explain the concept of \textit{brain lateralization}.
Secondly, we introduce an overview of research on the work of eyes and brain.
Lastly, we present activity monitoring works in the Human-Computer Interaction domain.

\subsection{Brain Lateralization}
\label{sec:brain_lateralization}
\textit{Brain lateralization} is an important concept that underpins our research. 
The concept is that the right and left hemispheres of the brain each have their responsibilities as cognitive functions~\cite{walker1980lateralization, rogers2021brain, rogers2024lateralization, ocklenburg2024lateralized}.

\citeauthor{toga2003mapping} work on an extensive review of structural and functional differences in left and right brain asymmetry~\cite{toga2003mapping}.
Their review mentions that one of the earliest observations of brain asymmetry is that the left hemisphere is used for ``language production,'' including grammatical processing, semantic knowledge, and syntax~\cite{dapretto1999form, binder2000new}.
Supporting their points, recent research led by \citeauthor{kong2018mapping} states that the left hemisphere of the brain has a large Broca's area, which is known as an important region for ``speech production''~\cite{kong2018mapping, neubauer2020evolution, kong2022mapping}.
The Broca's area was discovered due to the observation that patients with tumors and strokes in the left hemisphere result in severe language impairment~\cite{toga2003mapping}.
Another role of the left hemisphere is to process arithmetic~\cite{semenza2006math, funnell2007calculating, pinel2010beyond}.
\citeauthor{funnell2007calculating} work on the simple calculation experiment against two cerebral hemispheres of a split-brain patient~\cite{funnell2007calculating}.
As for the experiment, the patient conducted four arithmetic tasks.
The study confirmed that when the left hemisphere of the brain was operating normally, calculation performance was higher than in subjects with only the right hemisphere of the brain.
The work was then extended by \citeauthor{pinel2010beyond} by recruiting 209 healthy subjects to investigate how language processing (reading and listening) and arithmetic correlate with left brain hemisphere~\cite{pinel2010beyond}.
Both language processing and arithmetic observed lateral activation of the same two regions in the brain.

On the other hand, drawing is considered the right hemisphere's role~\cite{kirk1989hemispheric, nikolaenko2003artistic}.
\citeauthor{kirk1989hemispheric} work on comparing drawing tasks between 69 consecutive stroke patients with single cerebral lesions on CT, who were all right-handed.
Within 69 patients, 41 patients are right brain damaged, and 28 patients are left brain damaged.
The task was to draw a circle, cube, square, tree, house, and person.
The experiment confirmed that right brain damaged patients displayed hemispatial neglect and impaired spatial
relationships~\cite{kirk1989hemispheric}.
In the case of the left brain damaged subjects, their hand paralysis was severe because they were right-handed and could not perform well on drawing.
\citeauthor{nikolaenko2003artistic} states that artistic thinking, or so-called non-verbal visuo-spatial thinking, is supported by the right brain hemisphere~\cite{nikolaenko2003artistic}.
The research suggests that the right hemisphere's role is significant in the early stage of the creative process, in the conception stage.
Another role of the right brain hemisphere is processing music~\cite{gordon1983music, peretz2005brain}.
\citeauthor{gordon1983music} evaluated an experiment against a musician who got damage in the right occipitotemporal cortex~\cite{gordon1983music}.
The research discovered that musicians with right brain damage can write or read text but can no longer read musical notes by naming, singing, or playing them.
\citeauthor{peretz2005brain} states that musical pitch-based (melodic) and time-based (temporal) relations with the right brain hemisphere~\cite{peretz2005brain}.
This means listening to music and observing pitch and rhythm requires using the right brain hemisphere.

As mentioned above, the left brain hemisphere is involved in language processing and arithmetic, whereas the right brain hemisphere is involved in drawing and listening to music.

\subsection{Eyes and Brain}
This section describes the relationship between the eye and the brain.
We first explain existing eye-tracking technologies.
Then, we introduce how those eye-tracking technologies are used to discover the relationship between brain functions.

Eye-tracking methods include computer mounted device~\cite{grootjen2024uncovering}, webcam~\cite{zhang2019evaluation, Hisadome_2024_WACV}, and wearable~\cite{reddy2024towards, berkovsky2019detecting, putze2020platform}.
Computer-mounted devices include using Tobii software and the eye-tracker~\cite{grootjen2024uncovering}.
The computer-mounted Tobii eye-tracker is evaluated, and it is mentioned that the tracking is consistent through the display~\cite{housholder2021evaluating}.
Another approach is using webcam or so-called appearance-based estimation~\cite{zhang2019evaluation, Hisadome_2024_WACV}.
This approach aims to estimate the gaze direction from eye features and solve problems using computer vision.
Lastly, the wearable device includes a head-mounted display~\cite{reddy2024towards} or glasses~\cite{berkovsky2019detecting, putze2020platform}.
Sensors like infrared or RGB cameras are mounted on the wearable hardware.

Using these eye-tracking technologies, studies have explored the relationship between the eyes and the brain~\cite{london2013retina, krejtz2018eye, kucewicz2018pupil, oliva2018pupil, joshi2020pupil}.
\citeauthor{london2013retina} mention that the retina comprises layers of specialized neurons interconnected through synapses.
The retina is an extension of the central nervous system (CNS), and much of what is learned from eye research could apply to the brain and spinal cord, and vice versa~\cite{london2013retina}.
Among brain-related diseases, stroke, multiple sclerosis, Parkinson's disease, Alzheimer's disease, and lymphoma are associated with the retina.
Those symptoms influence the eye behaviors such as visual acuity, contrast sensitivity, and color blindness.
\citeauthor{kucewicz2018pupil} introduced the eyes as a biomarker of human memory performance.
The work investigated changes in pupil size during encoding and recall of word lists~\cite{kucewicz2018pupil}.
The study discovered that words that were successfully recalled showed significant differences in pupil response during their encoding phase compared to words that were forgotten. 
For successfully recalled words, the pupil size was more constricted before and more dilated after the onset of word presentation.
This result shows that pupil size or diameter significantly relates to the brain.
Lastly, \citeauthor{oliva2018pupil} works on discovering the change in pupil size fluctuations and the process of emotion
recognition~\cite{oliva2018pupil}. 
The experiment is conducted by preparing human nonverbal vocalizations (e.g., laughing, crying) and indicating the emotional state of the speakers as soon as they identify it.
The study discovered that during emotion recognition, the decision-making process drives the time course of pupil response.
In particular, peak pupil dilation betrayed the time of emotional selection. 
Moving into the direction of eye movements, the work of \citeauthor{canosa2009real} suggests that there are noticeable differences in fixation duration while performing numerous activities in daily life~\cite{canosa2009real}. 
These can lead to brain activation of specific neural signals. 
Also, fixation duration has been declared a useful metric to study learning processes~\cite{negi2020fixation}, which implies its effect on the brain.

These studies suggest that the pupil is related to brain state. Pupil size is an interesting parameter for understanding human cognitive states.

\subsection{Cognitive Monitoring in Human Computer Interaction}
Cognitive monitoring has been introduced in the human-computer interaction domain~\cite{yan2022emoglass, ning2023smartphone, watanabe2023engauge, gjoreski2023ocosense}.
\citeauthor{ning2023smartphone} use smartphone keyboard input and accelerometer to measure brain health~\cite{ning2023smartphone}.
The study recruited 85 clinical samples and discovered that the group with higher cognitive performance typed faster and was less sensitive to the time of day. 
The study aims to measure brain health through the daily use of smartphones.
\citeauthor{gjoreski2023ocosense} introduce OCOsense smart glasses to monitor daily facial gestures and expressions using wearable sensors~\cite{gjoreski2023ocosense}.
The hardware application is proposed as a tool for remote mental health monitoring.
EnGauge aims to measure facial gesture and expression, especially engagement during online communication~\cite{watanabe2023engauge}.
Instead of wearable sensing, the application uses a camera as a sensor to measure daily engagement.
\citeauthor{yan2022emoglass} introduce EmoGlass, a smart glasses monitoring device to capture seven emotions using a camera in real-time~\cite{yan2022emoglass}.
The Emotion Diary shows the transition of emotions as a diary and presents to the user.
The application supports users to self-monitor their daily emotions to lead to emotional health.

Our application aims in a similar direction to the one presented above.
As is introduced in Figure~\ref{fig:teasure}, we aim to make left and right brain lateralization activity accessible and also enable daily monitoring.


\begin{table}[t!]
    \centering
    \renewcommand{\arraystretch}{1.5}
    \caption{Summary of Experiments Performed in the Study. Four types of activities are included: two left-brain dominated activity like arithmetic and essay writing, and two right-brain dominated activity like drawing and listening to music.}
    \scalebox{0.9}{
    \begin{tabular}{cccl}
        \hline
        \textbf{Activity} & \textbf{Label} & \textbf{Brain Lateralization}  & \textbf{Evidence} \\
        \hline
        Arithmetic & L(math) & Left & \citet{semenza2006math}, \citet{pinel2010beyond} \\
        Drawing & R(drawing) & Right & \citet{kirk1989hemispheric}, \citet{nikolaenko2003artistic} \\
        Essay Writing & L(language) & Left & \citet{toga2003mapping}, \citet{neubauer2020evolution} \\
        Listening to Music & R(music) & Right & \citet{gordon1983music}, \citet{peretz2005brain}  \\
        \hline
    \end{tabular}
    }
    \label{tab:experiment_summary}
\end{table}

\section{Data Collection}
We used the Tobii Pro eye-tracker~\cite{TobiiProLab} with a sampling rate of 90Hz to record the data. Four tasks corresponding to each type of brain activity were designed, as described in Table~\ref{tab:experiment_summary}. These tasks were to be performed entirely on a computer having a screen resolution of $1280 \times 720$ pixels. A controlled environment was maintained for the experiments.
As the relative baseline size of the pupil changes throughout the day, the experiments were performed by all the participants at the same time during the day~\cite{daguet2019baseline}. 
A uniform luminance was maintained inside the laboratory to reduce the impact of illumination on the pupils~\cite{xu2011pupillary}. We captured eye recordings during each session of the experiment.

\subsection{Participants}
The experiments were performed by 34 individuals, 17 male and 17 female, aged between 22 and 32 years. 
The average age of the participants was 25 years. 
The participants were of South Asian, Southeast Asian, European, and South African origin. 
These were either university students or working professionals in Germany. 
We obtained the informed consent of all the participants according to General Data Protection Regulation (GDPR).
The participants were instructed about the procedure for data collection.
Every participant was asked to calibrate the eye tracking system at the beginning of the experiment. 
Participants were allowed to wear spectacles or contact lenses as per their preference in everyday life. They had the option to opt out of the experiment at any time. The experiment was performed by each individual in isolation in the lab to control the effects of external erroneous factors.

\begin{figure}[t!]
    \centering
    \scalebox{1.0}{\includegraphics[width=\linewidth]{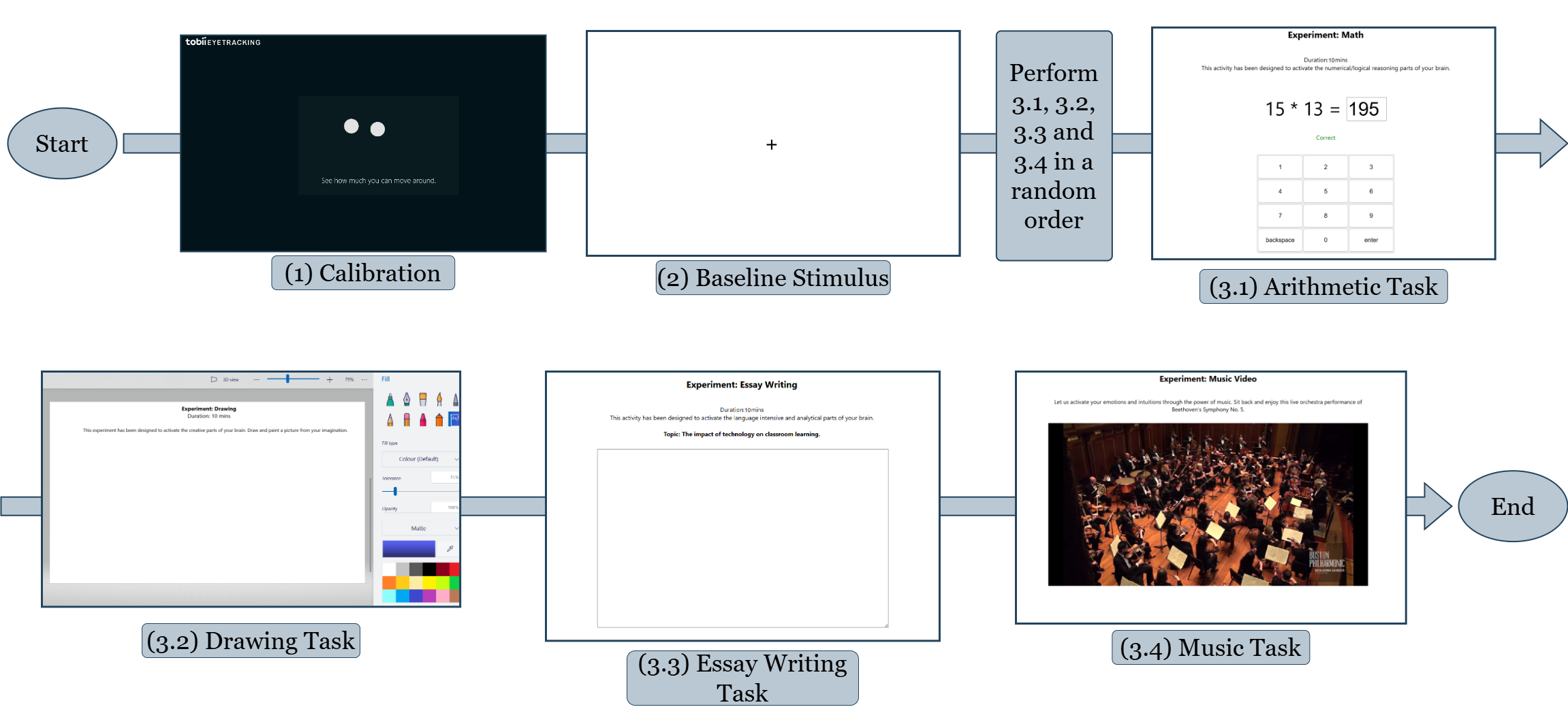}}
    \caption{Experiment flow. Procedure with step-wise screenshots of the computer during any session of data collection.}
    \label{fig:exp_setup}
\end{figure}

\subsection{Experiment Details}
The lateralization of brain functions is based on general trends experienced in healthy individuals.
As stated in Section~\ref{sec:brain_lateralization}, cognitive abilities like language processing, analytical thought, logic, memory, and problem-solving are predominantly controlled by regions on the left side of the brain~\cite{semenza2006math, pinel2010beyond, toga2003mapping, neubauer2020evolution}.
Others, like creativity, intuition, and emotional responses, are controlled by regions on the right side~\cite{kirk1989hemispheric, nikolaenko2003artistic, gordon1983music, peretz2005brain}.

We designed four tasks, two left-brain and two right-brain dominated activities, to explore the pupillary responses caused by neural processes from distinct sides of the brain. Table~\ref{tab:experiment_summary} summarizes the experiments performed for each cognitive activity. 
In the arithmetic task, beginner- and intermediate-level mental math problems, like subtraction and multiplication, were presented for the participant to solve.
The essay writing task required participants to evaluate and write about their opinions on a given topic. 
The topics included analytical writing subjects related to technology, psychology, and social media. 
In the drawing task, participants were asked to make a painting of their choice on the Paint 3D application interface. 
We placed no restrictions on the kind of picture to be drawn. 
It was beneficial for the participants to use their imagination to generate the corresponding pupillary eye movements. 
In the fourth task, participants were shown a music performance of an orchestra on YouTube with the intent of activating emotional cues within them.

\subsection{Data Acquisition Procedure}
In the study, every participant was required to perform the tasks mentioned above while the eye-tracker recorded their eye movements. The experimental procedure is shown in Figure~\ref{fig:exp_setup} with screenshots captured at each step.
Each participant was assigned tasks in a random order. 
This was done to reduce inter-task correlation, if any, and maintain variability. 
Hence, the true data class labels were decided by the order of the experiment. We built a simple web application to display these tasks on screen, automatically start and stop the data recordings, and maintain time. 
Each task takes ten minutes.
At the beginning of each task, a blank screen containing a single focal point stimuli was presented for 45 seconds. 
This constitutes the baseline period to capture the pupil diameter at a neutral stimulus, necessary for normalization~\cite{yau2021evidence, mathot2018safe}. 
The next screen displayed the activity task for the predefined duration. 
At the end of this duration, the display was disabled. These steps were repeated for all four task sessions.

\subsection{Dataset Preparation}
Our study used the default data produced by the Tobii software. 
This comprised a separate CSV file for every recorded session, containing raw data that the eye-tracker collects, such as timestamps, gaze positions, confidence levels, and pupil diameter. 
As mentioned before, the true data was labeled by deciding the experiment order for each participant. 
During implementation, each interim stage of data was prepared as a pickle file and stored for the next stage to avoid the usage of extra memory and computing resources. 
Additionally, we maintained participant feedback questionnaire responses.

\section{Methodology}
Our research aims to investigate the relationship between pupillometry and brain lateralization. 
We aim to put forward an efficient methodology for classification.
We employ a sliding window algorithm to extract meaningful features to train on classical machine-learning models and compare their performance. 
Features are then extracted from eye movements, such as left and right pupil diameter time series and fixations, and monitor the classification performance at the single activity level.
Finally, we explore a few conditions and parameters that have the potential to enhance the evaluation metrics for brain lateralization activity recognition.

\begin{figure}[t!]
    \centering
    \includegraphics[width=0.8\linewidth]{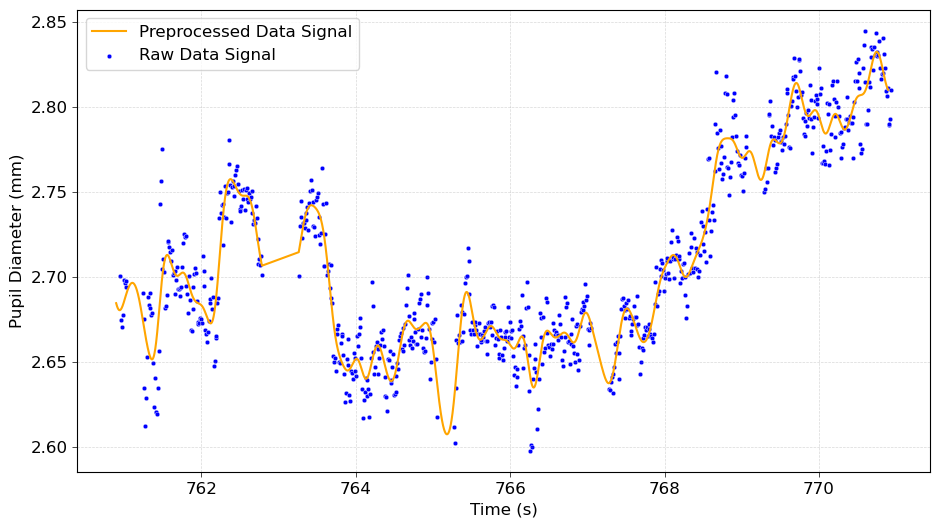}
    \caption{Variation of raw and preprocessed pupil diameter signal over time.}
    \label{fig:pp_raw_pd}
\end{figure}

\subsection{Preprocessing}
The raw eye-tracking data contains noisy or invalid samples, discontinuities, and outliers caused by external erroneous factors, like room brightness, eye obstruction, looking away from the screen for longer intervals.
Also, normalization of the pupil diameter is an essential step to handle participant-induced bias~\cite{aminihajibashi2019individual}.

We devised a pipeline to preprocess the recorded eye data by following the method and guidelines introduced by \citet{kret2019preprocessing}. 
First, the raw pupil diameter time series for both eyes was filtered to remove invalid samples. 
These included samples marked "invalid" by the eye-tracker, blank or null values, and incorrect values of pupil size. 
In the next step, we remove the possible noise and outliers. 
These included the following: 

\begin{itemize}
    \item Samples outside the median absolute deviation (mad) of pupil dilation speed.
    \item Samples deviating from the trend-line of the pupil signal.
    \item Samples around huge gaps in the data, also known as edge artifacts.
    \item Sparse or isolated samples, mainly caused by blinking.
\end{itemize}

The pupil diameter signal was then interpolated with a higher sampling rate of 1000 Hz, then smoothing using a zero-phase low-pass filter. Figure~\ref{fig:pp_raw_pd} visualizes the interpolated smooth signal generated at the end of the preprocessing pipeline overlayed on the raw signal.

Figure~\ref{fig:exp_setup} show the overall experiment flow.
The resulting samples were normalized through subtractive baseline correction~\cite{mathot2018safe}. 
We used the average pupil diameter values recorded during the period of no stimulus as the baseline pupil diameter.
Finally, the pupil diameter sequence obtained after baseline correction was used as input to the next steps.

\begin{table}[t!]
\centering
\caption{Statistical features extracted from the pupil diameter and fixation categories of eye movements.}
    \begin{tabularx}{\textwidth}{>{\arraybackslash}p{3cm} >{\arraybackslash}p{2.5cm} >{\arraybackslash}p{6.2cm} >{\arraybackslash}X >{\arraybackslash}X}
        \hline
        \textbf{Category} & \textbf{Sub-category} & \textbf{Feature Function} & \textbf{Type} & \textbf{Count} \\
        \hline
        & & & \\
        Pupil Diameter (PD) & Left PD, Right PD & mean value, standard deviation (std), median absolute deviation (mad), minimum value (min), maximum value (max), energy, signal magnitude area (sma), entropy, interquartile range (iqr) & T, F & T:18 F:18 \\
        & & & & \\
        \cline{3-5}
        & & & & \\
        & & range, root square means (rms), fourth order Burg auto-regression coefficient, pearson correlation & T & 7 \\
        & & & & \\
        \cline{3-5}
        & & & \\
        & & kurtosis, skewness, weighted average frequency component (meanFreq), largest frequency component (maxFreq), spectral energy of frequency band, power spectral density (psd) & F & 12 \\
        & & & & \\
        \hline
        & & & & \\
        Fixation (Fx) & Fx Duration & mean value, standard deviation (std), median absolute deviation (mad), minimum value (min), maximum value (max), energy, signal magnitude area (sma), entropy, interquartile range (iqr), range, root square means (rms) & T & 11 \\
        & & & & \\
        \cline{3-5}
        & & & \\        
        & Fx Count & frequency of occurrence of fixations (Fx\_count) & - & 1\\
        & & & & \\
        \hline
    \end{tabularx}
    \label{tab:features_list}
    \textit{\newline T: Time Domain, F: Frequency Domain} 
\end{table}

\subsection{Sliding Window and Feature Calculation}
\label{sec:sliding}
Previous work has shown that the sliding window method can improve the generalization of a machine-learning model operating on sequential data~\cite{tanaka2022sliding}. 
It can enable data analysis by sliding the window within a specified length with some overlap to retain contextual information of a time series sequence. Following a similar approach, we employed the sliding window algorithm to extract features from the preprocessed pupil diameter signal. 

Further, feeding heterogeneous data, a combination of implicit features extracted from raw data of the eye-tracker and explicit statistical features extracted from pupil data can lead to decent results in related problems~\cite{vortmann2021combining}.
\citet{watanabe2021discaas} utilized sixty statistical features extracted from time-series input from a camera to analyze the micro-behavior of participants. 
These included features in the time and frequency domain. Inspired by this, we expanded a similar concept to extract statistical features from eye movements. Specifically, we focused our investigation on pupil diameter and fixation duration components. 
Table~\ref{tab:features_list} shows the features list used in this study.

We explored how varying the window size in a certain activity duration would impact brain lateralized cognitive activity. We employed an empirical methodology to find the optimal window size and activity duration. First, we analyzed different values of activity duration, from two minutes to eight minutes. 
Previous work has shown the consideration of smaller windows in sliding window algorithm~\cite{ishimaru2014blink, tanaka2022sliding}. 
In our work, we wanted to attain a robust estimation method for a longer time interval. The parameter values for the window size were 5s, 15s, 30s, 45s, 60s, 75s, 90s, and 105s. 
We maintained an overlap of 50\% of the window size. 
We calculated the features from Table~\ref{tab:features_list} for every sliding window. These were then provided as input to the machine-learning models.

\subsubsection{Extracting Temporal and Spectral Statistical Features from Pupil Diameter Time Series}
While frequency domain features can help understand power distribution across frequency components, time domain features can capture trends and complex patterns over time. 
We expected these features to provide a meaningful data representation when put together. We extracted twenty-five time-domain features. 
In this type, the temporal properties were derived by directly computing the statistical formulation of the pupil diameter (PD) time series. 
We computed the discrete Fourier Transform (FFT) of the spectral features of the PD data. 
Thirty frequency domain features were calculated from the FFT of the spectrum. 
As left and right PD signals could influence each other, we also included the Pearson Correlation Coefficient as a feature. Each of these features was calculated in the sliding window. 
Every run of the sliding window operated as one row of feature input.  

\subsubsection{Extracting Features from Fixation Eye Movements Category}
Eye-tracking generally covers a wide range of features extracted from two primary eye movements: saccades and fixations. While saccades contribute more to building a mental map of the scene, fixations obtain information on a particular focal point. 
We expect that saccades will have little relevance to our research because our goal is to focus on neural function processing through pupillometry rather than the visual perception of the eye.
Thus, we did not consider saccades in our study. 
A similar reasoning applies to widely used fixation features like fixation gaze points and fixation dispersion area.
 
We hypothesized that fixations, in conjunction with variation in pupil diameter, can influence brain cognitive activity. Thus, we were interested in analyzing differences in brain functions while including and excluding fixations. How long the fixation occurs could also be substantial in identifying neural functionality. For instance, the essay writing task would require a longer fixation while focusing on significant words of the sentence, compared to solving an arithmetic question that depends on memory and mental calculation.
Thus, fixation duration would be essential for our study. 
 
We calculated eye fixations in every sliding window interval. The fixation duration was derived from each fixation. Then, temporal statistical features were computed from the series of fixation duration values in the time window, Table~\ref{tab:features_list}. A total of eleven time-domain features were obtained. We also included the frequency of fixations during the sliding window interval, or $fx\_count$, as a feature.

\subsection{Classification}
Our dataset consists of 136 distinct time series, with 34 participants and four brain cognitive activities, making it a time series classification problem. 
As mentioned above, two out of four tasks belong to each type of brain hemispherical dominance. Additionally, our primary goal is brain lateralization recognition. Hence, this was treated as a binary classification problem with true labels: ``left'' assigned to arithmetic and essay writing tasks, and ``right'' assigned to drawing and music listening tasks.

The extracted features were fed to classical machine-learning models for classification. We chose SVM with an ``rbf'' kernel and default hyper-parameter setting to derive a baseline performance on binary classification. 
The presence of high non-linearity of features was expected to make inaccurate predictions with the SVM classifier.
We also used ensemble learning models like XGBoost and Random Forest, which are known to work well with multidimensional datasets. 
These can decrease variance through bagging and bias through boosting. 
They also have the advantage of parallelized ensemble learning. 
For the XGBoost classifier, we used specific hyper-parameters to guide the model's learning process: 100 estimators and the logloss evaluation metric. 
The hyper-parameter used in Random Forest was the number of trees chosen with a value of 100.

Finally, we wanted to investigate the prediction of individual brain cognitive functions. In this case, we performed a four-label classification with true data labels: L(math), L(language), R(drawing), and R(music) using the best-performing classifier from binary classification.

\subsection{Performance Evaluation}
\subsubsection{Feature Extraction Method Evaluation}
We evaluated our feature extraction-based method through leave-one-participant-out cross-validation (LOPOCV). 
This was used to determine how cognitive activity can be recognized irrespective of participant bias. 
The dataset was split so that one participant's data was used as test data in each iteration, and the remaining participants' data was used as training data. 
These steps were repeated until all participants were used for testing. 
We compared the performance of LOPOCV in two scenarios: using pupil diameter feature extraction with and without fixation category features. 
We also employed the best-performing classifier to deduce importance scores for the features.

\subsubsection{Analysis of Contributing Factors}
We explored the impact of window size and average fixation duration in a specific window on brain lateralized activity. We monitored performance evaluation for various parameter values for these conditions, as mentioned in Section~\ref{sec:sliding}. 
The evaluation metric used was the F1 score. 
We also utilized the best-performing classification model to measure the contribution of average fixation duration threshold values.

\subsubsection{Individual Activity Performance}
We implemented the leave-one-activity-out cross-validation (LOAOCV) approach. First, we divided the data into ten folds. Within each fold, the following steps of LOAOCV were implemented.
As the name suggests, we utilized the data corresponding to the left-out activity label for testing and the remaining data for training. This was repeated for all four activities in new iterations. In the case of high-class imbalance, we used the synthetic minority over-sampling technique (SMOTE). In the end, the predictions of LOAOCV in each fold were combined to obtain the overall classification rate for each brain activity.

\begin{figure}[t!]
    \centering
    \includegraphics[width=0.6\textwidth]{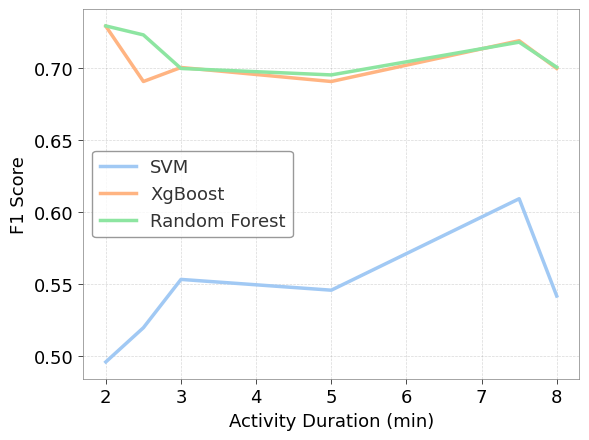}
    \caption{Variation in weighted average F1 score with activity duration for machine-learning models.}
    \label{fig:f1_act_fig}
\end{figure}

\section{Results}
This section presents the performance evaluation of machine-learning models in determining left-right brain lateralization. 
First, we determine what duration of the activity data is sufficient to capture patterns and provide good insights. 
Then, we explain the influence of the sliding window's length by comparing the evaluation metrics for different window sizes. 
This is done for two types of features using leave-one-participant-out cross-validation:

\begin{itemize}
    \item \textit{PD}: Temporal and spectral features extracted from pupil diameter.
    \item \textit{PD + Fx}: Combination of PD features with features extracted from fixations, as given in Table~\ref{tab:features_list}.
\end{itemize}

We investigate the impact of fixation duration. 
We compare model performance by selecting varying threshold ranges of fixation duration and feeding corresponding features to the model. 
Further explore how an individual activity can be predicted through leave-one-activity-out cross-validation. 
Lastly, we present and discuss the highest features of importance in recognition.

\subsection{Impact of Activity Duration}
\label{sec:impact}
We performed empirical analysis to choose an appropriate total activity duration for predicting brain lateralization. 
We plotted the F1 score over varying activity duration displayed in Figure~\ref{fig:f1_act_fig} to assess evaluation. 
The baseline period to measure the pupil diameter was forty seconds. 
We considered a minimum duration of more than twice the baseline to explore some noticeable trends in pupillometric responses. 
Thus, two minutes or more was chosen. 
An activity duration between 2-3 minutes and 6-8 minutes gave a high evaluation result for XGBoost and Random Forest. 
The latter interval had the best-observed result for SVM. 
The F1 score of the three classifiers showed a similar upward trend and peaked at 7.5 minutes. 
A large activity duration corresponds to the presence of enough training data and well-represented input features for the time series data. 
We concluded that a duration of 7.5 minutes could be chosen for further performance evaluation, which formed a good preliminary condition for our analysis.

\subsection{Impact of Window Size}

\begin{table}[t!]
\renewcommand{\arraystretch}{1.2}
\centering
\caption{Comparison of F1 score for leave-one-participant-out cross-validation with variable window sizes for two types of feature sets: PD and PD+Fx.}
\begin{tabularx}{\textwidth}{c|X X X X X X}
    \hline
    \multirow{2}{*}{\textbf{Window Size (s)}} & \multicolumn{2}{l}{\textbf{SVM}} & \multicolumn{2}{l}{\textbf{XGBoost}} & \multicolumn{2}{l}{\textbf{Random Forest}} \\
    & \textit{PD} & \textit{PD + Fx} & \textit{PD} & \textit{PD + Fx} & \textit{PD} & \textit{PD + Fx} \\
    \hline
    5 & 0.542 & 0.581 & 0.699 & 0.772 & 0.705 & 0.775 \\
    15 & 0.545 & 0.633 & 0.714 & 0.768 & 0.720 & 0.770 \\
    30 & 0.546 & 0.643 & 0.721 & 0.790 & 0.723 & 0.784 \\
    45 & 0.555 & 0.656 & 0.735 & 0.781 & 0.726 & 0.776 \\
    60 & 0.568 & 0.653 & 0.744 & 0.797 & 0.739 & 0.789 \\
    75 & 0.546 & \textbf{0.679} & 0.734 & 0.778 & 0.724 & 0.778 \\
    90 & 0.550 & 0.655 & 0.739 & \textbf{0.813} & 0.749 & \textbf{0.810} \\
    105 & 0.582 & 0.662 & 0.725 & 0.806 & 0.736 & 0.798 \\
    \hline
\end{tabularx}
\label{tab:lopocv}
\end{table}

\begin{figure}[t!]
    \centering
    \begin{subfigure}[b]{0.25\textwidth}
        \centering
        \includegraphics[width=\textwidth]{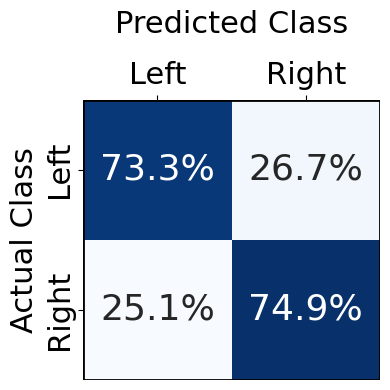}
        \caption{PD}
        \label{fig:conf_matrix3}
    \end{subfigure}%
    \hspace*{0.75cm}
    \begin{subfigure}[b]{0.25\textwidth}
        \centering
        \includegraphics[width=\textwidth]{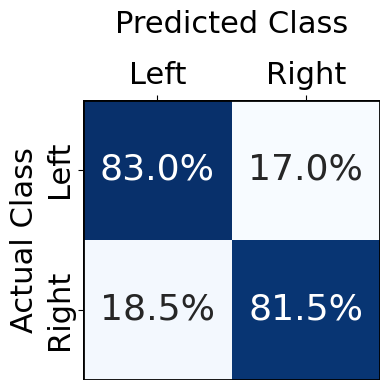}
        \caption{PD+Fx}
        \label{fig:conf_matrix4}
    \end{subfigure}
    \caption{Confusion matrix of leave-one-participant-out cross-validation (LOPOCV) for the best performing window size of 90 seconds in XGBoost.}
    \label{fig:conf_matrices}
\end{figure}

Table~\ref{tab:lopocv} shows the results of incrementally increasing window lengths for two types of feature input, \textit{PD} and \textit{PD+Fx}, by evaluating the performance of three classifiers - SVM, XGBoost and Random Forest using leave-one-participant-out cross-validation. A 50\% overlap was maintained for each sliding window. All the results were compiled for an activity duration of 7.5 minutes. 

There was an evident rise in F1 score on utilizing the \textit{PD+Fx} features for every window size. In almost every recording, there was an average improvement of more than 5\% in the F1 score. 
Another observation was that the larger the window size, the higher the F1 score until the point of maximum was reached. 
The highest F1 scores, 0.813 and 0.810, were achieved by XGBoost and Random Forest classifiers, respectively, for a window size of 90 seconds. 
To further verify this result, Figure~\ref{fig:conf_matrices} shows a confusion matrix of XGBoost. 
We could verify that the brain lateralization recognition rate improved and the error rate reduced after adding fixation features. 
In the case of SVM, 75 seconds is the most optimal window size. 
The range of F1 score values was comparably low for SVM, probably due to the multidimensional feature input. 
Although there is no optimal performing window size, we can roughly estimate that a window greater than 30 seconds would be most advantageous.

\subsection{Impact of Fixation Duration}
In the previous section, we concluded that adding fixation features greatly improved our predictions. 
Here, we explored the influence of fixation duration in classification. 
We filtered out that \textit{PD+Fx} features where the average fixation duration of the sliding window was within the specified threshold and then used them for classification. 
Due to limiting the feature set, some threshold intervals did not have enough training data and could not give classification predictions. 
The F1 scores of the XGBoost classifier are shown in Table~\ref{tab:fix_dur}. We used sliding window sizes 5s, 45s, and 90s to understand the relationship between window length and fixation duration. 

We observed that the F1 score increased rapidly as the fixation duration increased, reaching a maximum value at a fixation duration between 500 and 750 ms, followed by a decline. 
For the 90s window, features with an average fixation duration within 250 - 1000 ms contributed predominantly to activity estimation. 
This is seen from the higher F1 scores, 0.859, 0.894, and 0.855, and a notable drop after the fixation duration exceeds 2 seconds. 
The observed behavior for the 5s and 45s sliding windows is slightly different. 
Although there is an F1 score maximum at a fixation duration interval of 500 - 750 ms, the performance gap among the threshold ranges declined gradually. 
While all fixation duration ranges contribute somewhat to activity estimation, a shorter fixation duration is preferable for any window size. 

\begin{table}[t!]
\renewcommand{\arraystretch}{1.2}
\caption{Variation of leave-one-participant-out cross-validation F1 score over threshold ranges of average fixation duration for multiple window sizes.}
\begin{tabular}{c|ccc}
    \hline
    Average Fixation & PD + Fx & PD + Fx & PD + Fx \\
    Duration Threshold (ms) & (sliding window = 5s) & (sliding window = 45s) & (sliding window = 90s) \\
    \hline
    < 250 & 0.611 & 0.333 & - \\
    250 - 500 & 0.816 & 0.834 & 0.859 \\
    500 - 750 & \textbf{0.824} & \textbf{0.890} & \textbf{0.894} \\
    750 - 1000 & 0.754 & 0.819 & 0.855 \\
    1000 - 1500 & 0.785 & 0.818 & 0.754 \\
    1500 - 2000 & 0.775 & 0.616 & 0.764 \\
    2000 - 2700 & 0.714 & 0.789 & 0.533 \\
    2700 - 3500 & 0.703 & 0.777 & 0.333 \\
    > 3500 & - & 0.461 & 0.533 \\
    \hline
\end{tabular}
\label{tab:fix_dur}
\end{table}


\subsection{Impact of Cognitive Activity}

\begin{figure}[t!]
    \centering
    \begin{subfigure}[b]{0.33\textwidth}
        \centering
        \includegraphics[width=\textwidth]{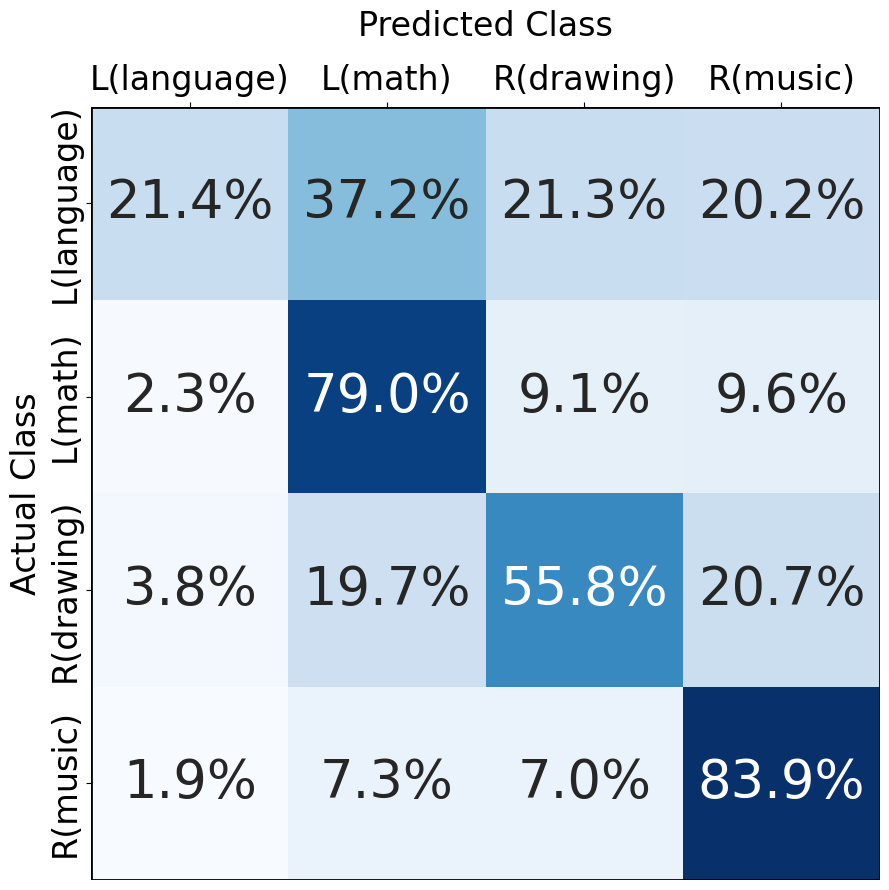}
        \caption{Sliding window = 5 s}
        \label{fig:cm_loaocv5}
    \end{subfigure}
    \begin{subfigure}[b]{0.33\textwidth}
        \centering
        \includegraphics[width=\textwidth]{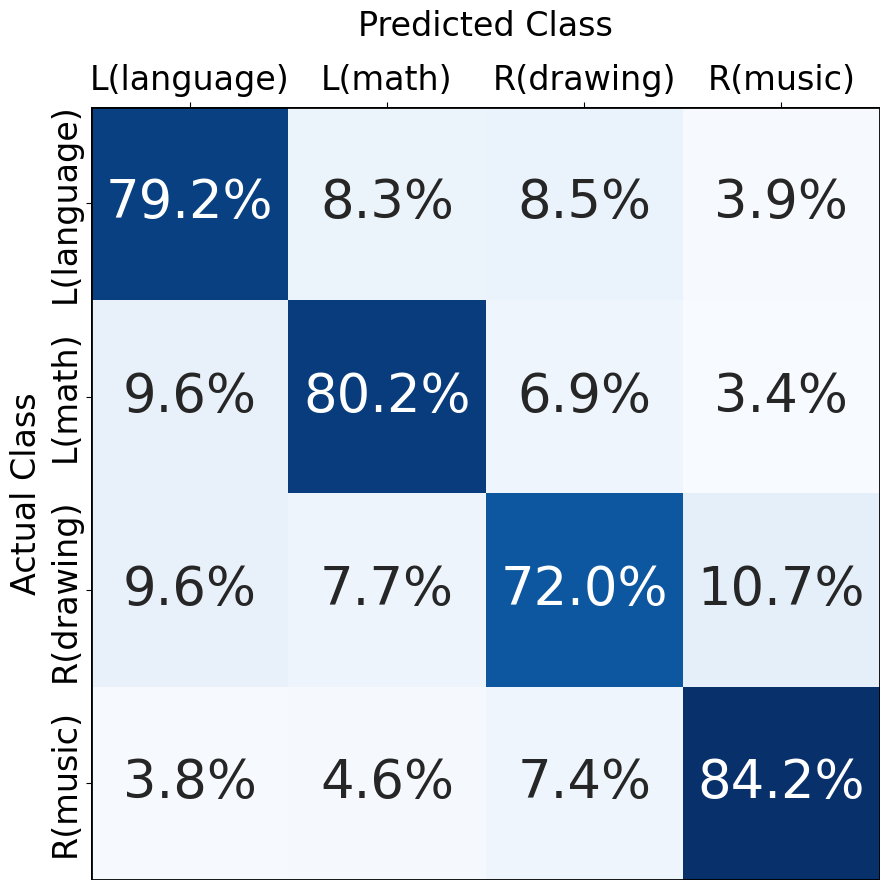}
        \caption{Sliding window = 45 s}
        \label{fig:cm_loaocv45}
    \end{subfigure}
    \begin{subfigure}[b]{0.33\textwidth}
        \centering
        \includegraphics[width=\textwidth]{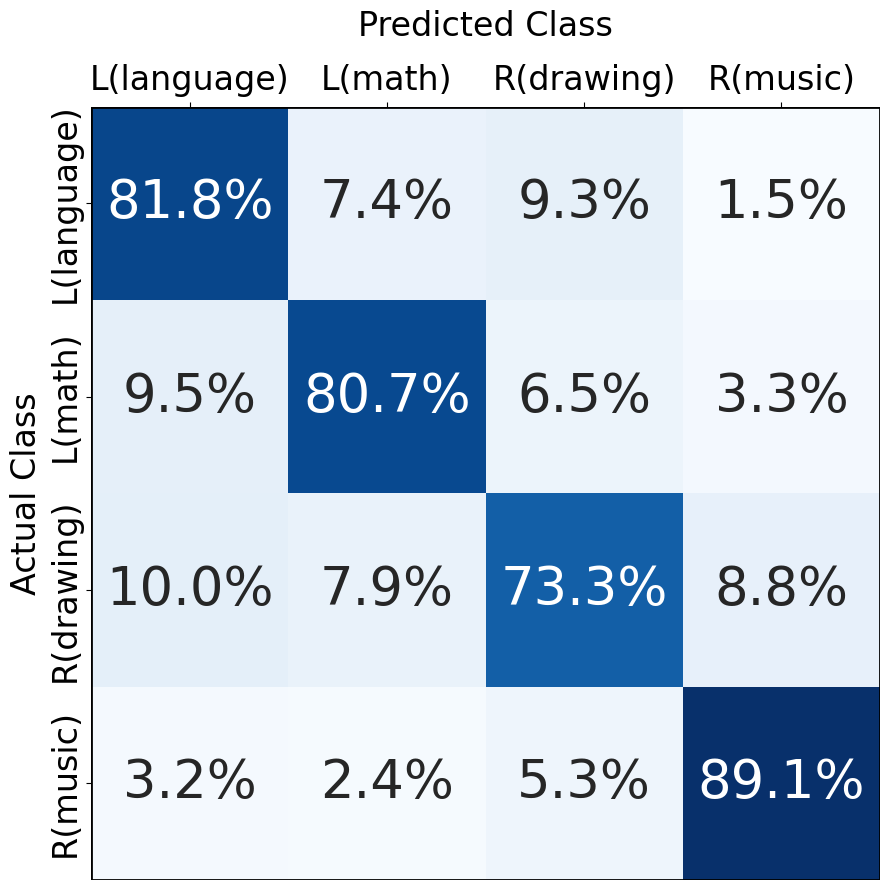}
        \caption{Sliding window = 90 s}
        \label{fig:cm_loaocv90}
    \end{subfigure}
    \caption{Confusion matrix of leave-one-activity-out cross-validation (LOAOCV) for three window sizes visualized for XGBoost classifier. Activities L and R denote left and right brain cognitive activities.}
    \label{fig:loaocv_cm}
\end{figure}




This study examines how well each cognitive activity could be estimated with \textit{PD+Fx} features. 
Figure~\ref{fig:loaocv_cm} shows the confusion matrix for leave-one-activity-out cross-validation to measure the classification rate for the four activities. 
The results confirm that the larger the window size, the better the classification result. 
R(music) had the highest recognition rate for all window sizes. This activity scored the highest classification rate, 89.1\%, out of all activities for the 90s sliding window. L(math) consistently performed in all cases, irrespective of the window size, with an average classification rate of over 79\%. 
We observed that activities L(language) and R(drawing) were recognized fairly well for larger windows and poorly for shorter windows. 
The L(language) activity was greatly misclassified as L(math), with a classification rate of 37.2\%. 
These two tasks had much higher error rates than L(math) and R(music). Section~\ref{sec:discussion} explains the reasons for poor performance. 
We can conclude that at the window size, in addition to activity type, plays a vital role in estimation.

\subsection{Feature Importance}

\begin{figure}[t!]
    \centering
    \includegraphics[width=0.95\textwidth]{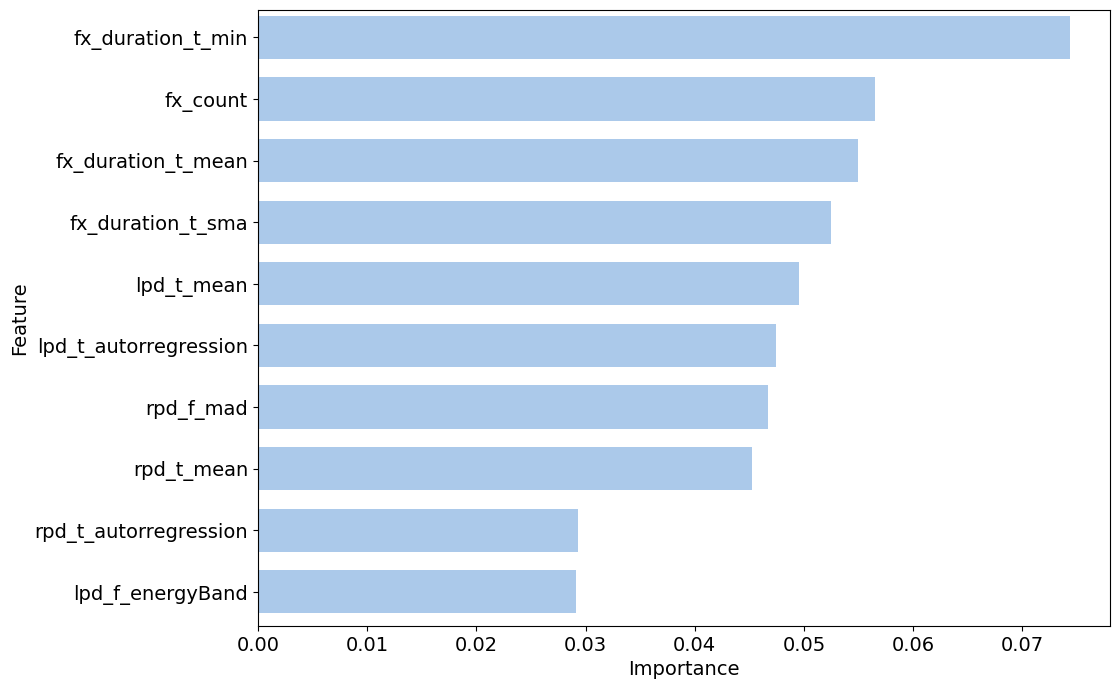}
    \caption{Top ten most important features according to XGBoost classification score.}
    \label{fig:feat_imp}
\end{figure}

We present the most influential features by calculating the XGBoost feature importance scores. 
Figure~\ref{fig:feat_imp} shows the top ten most important features.  
The four highest-ranked important features were fixation-based. These are the features with the prefix $fx$. Fixation count and three features extracted from fixation duration, minimum, mean, and signal magnitude area (SMA) constitute more than 5\% importance each. The highest importance value is assigned to  $fx\_duration\_t\_min$ with a value of 0.075, which is considerably higher than other features. We further discuss the high influence of this feature in Section \ref{sec:discussion}. The improvement in classification rate in using \textit{PD+Fx} features was also verified through this plot.
The highest-scoring time-domain features included mean, auto-regression coefficient, and median absolute deviation (mad).
These are denoted by the prefixes $rpd\_t$ and $lpd\_t$. 
Their scores were in the range of 0.03 to 0.051. 
Lastly, $lpd\_f\_energyBand$ is one of the significant features discovered from the frequency domain features.


\section{Discussion}
\label{sec:discussion}
In this section, we discuss how this work contributes to the direction of our research questions. We explain and evaluate the most important findings from our results.

\subsection{Effective Recognition of Brain Lateralization}
A major contribution of our work is the high brain lateralization recognition rate achieved for a couple of machine-learning models, specifically XGBoost and Random Forest. 
The results recorded for these classifiers proved that these were the appropriate models to handle time series data in our study. 
The LOPOCV method achieved a high F1 score of 0.813 for XGBoost. Additionally, we observed that XGBoost and Random Forest consistently performed well, and the F1 score was measured at over 0.76 in all conditions. The confusion matrix, Figure~\ref{fig:conf_matrices}, further supports our findings, where both classes had good estimation levels. 
These results also confirmed the effectiveness of the preprocessing pipeline. 
This pipeline included multiple filtering steps to remove noisy samples, which were adept at estimating participant-independent activity.

Previous work has shown that features related to saccades, like saccade length and angle, blink rate, gaze positions, gaze paths, fixation points, and many more, are among the most important features in eye tracking studies for similar use cases~\cite{lim2022eye, liversedge2000saccadic, bulling2009eye, vortmann2021combining}. 
Related work in desktop activity recognition has shown noteworthy results when a combination of regular low-level features (fixations and saccades) and abstract features calculated from low-level features (fixation dispersion area and saccade direction) were utilized~\cite{srivastava2018combining}. 
We used a methodology utilizing statistical pupillometric and fixation duration features that proved to be quite impactful in the predictions. 
We hypothesized that the change in right and left pupil diameter would be different for left and right brain activities.
Obtaining detailed pupil diameter features in the time and frequency domain was able to capture the specificity of these patterns. 
Moreover, the idea of including just fixation duration as a feature worked well because of this feature's high variability level in various cognitive processes~\cite{negi2020fixation}. 
Referring to the feature importance graph in Figure~\ref{fig:feat_imp}, the minimum and mean values of fixation duration and fixation count have the highest importance scores, each measuring above 5.5\%. 
Simple features like mean, autoregression coefficient, and median absolute deviation are other salient features. 
As a result, we established a simple and efficient feature extraction method that performed well despite lacking several widely used eye movement features.

\subsection{Identification of Favourable Conditions for Cognitive Activity Prediction}
Another contribution of this study was that we uncovered specific conditions for cognitive activity classification. We discovered that a longer window length is more optimal for prediction. The evidence supports this in Table~\ref{tab:lopocv} and Figure~\ref{fig:conf_matrices}.
Generally, a sliding window size of over 30 seconds is preferable. Although in our analysis, the highest-performing window size was 90 seconds, this parameter would still depend on the type of data and specific use case. 
Window size is also related to activity duration. For a long activity interval like ours, of 7.5 minutes, it was more efficient to acquire meaningful information in larger window segments.

Secondly, we learned that a subset of features within a specific range of average fixation duration contributed to brain lateralization estimation with a gap of more than 10\%. Table~\ref{tab:fix_dur} shows that this value is between 250 to 1000 ms. It is also definitive that window size influences fixation duration. 
We can observe more fine-grained prediction and less generalized information influence in smaller window sizes. 
This is why the short and medium fixation duration contribution levels were comparable in the 5s sliding window. 
The F1 scores obtained from 250 to 2000 ms verify this behavior.
Conversely, the prediction is more robust and generalized for larger window sizes. 
This resulted in greater dips in the F1 score from short to long fixation duration. 
Similar observations were shared in the work of \citet{srivastava2018combining}. 
These results benefit similar research topics based on the goal of the prediction.

Lastly, we discovered the importance of fixation eye movements in our use case. 
The minimum value of fixation duration in a sequence window had an importance score of 0.075, Figure~\ref{fig:feat_imp}. 
The model could distinguish the brain lateralization task based on the variation of this feature throughout the activity. 
We were restricting fixations in sliding windows, which led to the formation of an identifiable fixation occurrence rate. 
Thus, fixation count was another vital feature. The feature importance graph provided a deep analysis for RQ2. 

\subsection{Individual Activity Estimation}
\label{sec:discussion_3}
One of the core contributions of this study was to check how well any of the selected cognitive activities are recognized just through eye movements. Our feature extraction method was compelling enough to predict individual activities given certain preconditions, Figure~\ref{fig:loaocv_cm}. The music and arithmetic activities were constantly estimated well. Higher precision of eye-tracking data could have been collected with continuous attention on the screen for these tasks. 
Emotional cues induced by music cause dominant right-brain neural processes. The math brain functions instantly, resulting in memory signals and logical thinking, which are functions of the left brain. 
This led to a high classification rate.

There was a disparity in the results for language and drawing activity. We suspect that the eye-tracker must have recorded many obstructions while the participant performed the drawing task. Also, the distance between the eye and the screen must have been compromised in multiple instances while drawing on the screen. 
This could have led to the presence of more invalid and noisy samples. Research suggests that while artistic neural functions are predominantly right-brain dominated, the left hemisphere also plays a role. 
\citet{nikolaenko2003artistic} explained that the right brain generates nonverbal visual-spatial thinking, followed by an activation transfer from the right to the left brain in artistic thinking. 
This is a potential reason for the observed results in the drawing task. Additionally, the continuous movement of the participant's heads while typing and looking at the keyboard could have led to a higher error rate in the essay writing task. 
With the shorter window size of five seconds, the focus is on recognition in smaller intervals rather than generalization of the activity, which may also affect the results.

\section{Limitations and Future Work}
In this section, we will discuss the limitations of our work that can lead to potential future work.

The first limitation is that the cognitive tasks were designed to suit a more controlled environment due to the kind of eye-tracking technology used. As we used Tobii eye tracking~\cite{TobiiProLab}, we had to select experiments with desktop activities. We found that some activities had low individual classification rates. The reasons have already been discussed in Section~\ref{sec:discussion_3}. 
A possible future direction could be the exploration of new and upcoming eye-trackers like Invisible glasses or Core headband by Pupil Labs~\cite{PupilLabsCore, PupilLabsInvisible}. 
This would reduce the user's restriction in movement and provide the possibility of incorporating in-the-wild brain activities. 

The usability of such portable eye trackers could open up possibilities for numerous research works in real-time brain lateralization activity classification. It can pave the way for the \textit{EyeBrain} concept, introduced in Section~\ref{sec:intro} and its real world applications. 
This is one of the biggest motivations for our work. 
We are developing a real-time application that monitors brain lateralization activities during the day and provides a life log of daily statistics.
This could enable users to balance activities, focus on new activities to maintain their brain health, and help discover future life goals. 
This concept can support a variety of professions.

If real-time brain monitoring is considered, our study has another limitation: a long activity duration. 
We discovered that a longer activity duration is preferable for prediction, as per Section~\ref{sec:impact}. 
A shortcoming of this is the selection of a longer window size.
However, recording more than seven minutes for a single activity would be rather time-consuming. 
Real-time applications need to process window sizes larger than a few seconds and provide instant classification results.
Also, the pupillary movements specific to brain lateralization are expected to be observed right from where the eye receives the stimulus. Future work can be done on discovering an optimal minimum activity duration, preferably in units of a few seconds, that could process and recognize brain activity in real time.

The next limitation is that we utilized only classical machine-learning models for this study. 
Deep learning can handle more complex data and has the potential for high precision.
\citet{ismail2019deep} provided a literature review on recent deep neural network architectures used for time series classification. 
Future research can employ deep-learning-based brain lateralization estimation. Another approach could involve directly feeding time series sequential data rather than feature-extraction-based input. 
Models like long short-term memory models (LSTM), and transformers have outperformed classical machine-learning and traditional deep-learning models in processing sequential data~\cite{quddus2021using, sharma2021machine, stein2022eye}.

Another research direction could be improving our classification rate by adding a nuanced set of features. Improving the predictions of our feature extraction-based method is still possible. This could be done through the addition of multi-modal ocular information. One approach could be to record videos of the participants and utilize eye image frames at time steps. Employing engagement detection mechanisms, as in the work of\citeauthor{watanabe2023engauge}~\cite{watanabe2023engauge}, applying them to extract eye movements from the highest engaged time steps could be a potential idea too. Another approach could be incorporating EEG signal data, as has been done in other similar problems~\cite{rozado2015combining}.

Finally, our research still needs to integrate some phenomena like reverse brain lateralization and brain asymmetry. 
According to the work of \citet{morita2020right}, there is a possibility that left-handedness in certain individuals can alter the brain dominance of some cognitive functions of the right hemisphere. 
Similarly, there could be deviations from brain hemispheric cognitive specialization associated with evolutionary and genetic factors~\cite{corballis2014left}. 
These use-cases could lead to potential future work. Analyzing pupillometric data of left-handed individuals could be one possible approach. Another idea could be exploring left- and right-handedness as an additional feature of our method.

\section{Conclusion}
In this paper, we examined the ability of ocular metrics such as pupil diameter and fixation duration to recognize brain-lateralized cognitive functions. 
We thoroughly analyzed eye data from two left-brain-dominant activities, arithmetic and essay writing, and two right-brain-dominant activities, drawing and listening to music. 
Our methodology utilized two feature inputs: one with statistical temporal and spectral pupil diameter features and another combining the former with features derived from fixation duration. 
Our findings suggest that considering fixation duration is salient to the higher performance of machine-learning models. 
We also discovered certain parameters that positively influence brain lateralization classification, such as the sliding window size, the threshold range of fixation duration, and the most influential features. 
We obtained a high classification performance, with an F1 score of 0.894. 
Overall, our work achieved the primary goal of the \textit{EyeBrain} concept.

\begin{acks}
Claude Opus 4.7 supported the writing of this manuscript. Human check and revision were performed after each generation.
\end{acks}


\bibliographystyle{ACM-Reference-Format}
\bibliography{main}

@article{murphy2014pupil,
  title     = {Pupil diameter covaries with BOLD activity in human locus coeruleus},
  author    = {Murphy, Peter R and O'connell, Redmond G and O'sullivan, Michael and Robertson, Ian H and Balsters, Joshua H},
  journal   = {Human brain mapping},
  volume    = {35},
  number    = {8},
  pages     = {4140--4154},
  year      = {2014},
  publisher = {Wiley Online Library}
}

@article{joshi2016relationships,
  title     = {Relationships between pupil diameter and neuronal activity in the locus coeruleus, colliculi, and cingulate cortex},
  author    = {Joshi, Siddhartha and Li, Yin and Kalwani, Rishi M and Gold, Joshua I},
  journal   = {Neuron},
  volume    = {89},
  number    = {1},
  pages     = {221--234},
  year      = {2016},
  publisher = {Elsevier}
}

@article{london2013retina,
  title     = {The retina as a window to the brain—from eye research to CNS disorders},
  author    = {London, Anat and Benhar, Inbal and Schwartz, Michal},
  journal   = {Nature Reviews Neurology},
  volume    = {9},
  number    = {1},
  pages     = {44--53},
  year      = {2013},
  publisher = {Nature Publishing Group UK London}
}

@misc{TobiiProLab,
  author       = {Tobii AB},
  title        = {Specifications for the Tobii Eye Tracker 4C},
  howpublished = {\url{https://help.tobii.com/hc/en-us/articles/213414285-Specifications-for-the-Tobii-Eye-Tracker-4C}},
  address      = {Danderyd, Stockholm},
  year         = {2024}
}

@article{canosa2009real,
  title={Real-world vision: Selective perception and task},
  author={Canosa, Roxanne L},
  journal={ACM Transactions on Applied Perception (TAP)},
  volume={6},
  number={2},
  pages={1--34},
  year={2009},
  publisher={ACM New York, NY, USA}
}

@article{negi2020fixation,
  title={Fixation duration and the learning process: An eye tracking study with subtitled videos},
  author={Negi, Shivsevak and Mitra, Ritayan},
  journal={Journal of Eye Movement Research},
  volume={13},
  number={6},
  year={2020},
  publisher={European Group for Eye Movement Research}
}

@article{ismail2019deep,
  title={Deep learning for time series classification: a review},
  author={Ismail Fawaz, Hassan and Forestier, Germain and Weber, Jonathan and Idoumghar, Lhassane and Muller, Pierre-Alain},
  journal={Data mining and knowledge discovery},
  volume={33},
  number={4},
  pages={917--963},
  year={2019},
  publisher={Springer}
}

@article{sharma2021machine,
  title={Machine learning-based analysis of operator pupillary response to assess cognitive workload in clinical ultrasound imaging},
  author={Sharma, Harshita and Drukker, Lior and Papageorghiou, Aris T and Noble, J Alison},
  journal={Computers in biology and medicine},
  volume={135},
  pages={104589},
  year={2021},
  publisher={Elsevier}
}

@article{quddus2021using,
  title={Using long short term memory and convolutional neural networks for driver drowsiness detection},
  author={Quddus, Azhar and Zandi, Ali Shahidi and Prest, Laura and Comeau, Felix JE},
  journal={Accident Analysis \& Prevention},
  volume={156},
  pages={106107},
  year={2021},
  publisher={Elsevier}
}

@inproceedings{stein2022eye,
  title={Eye tracking-based lstm for locomotion prediction in vr},
  author={Stein, Niklas and Bremer, Gianni and Lappe, Markus},
  booktitle={2022 IEEE conference on virtual reality and 3D user interfaces (VR)},
  pages={493--503},
  year={2022},
  organization={IEEE}
}

@Article{watanabe2021discaas,
AUTHOR = {Watanabe, Ko and Soneda, Yusuke and Matsuda, Yuki and Nakamura, Yugo and Arakawa, Yutaka and Dengel, Andreas and Ishimaru, Shoya},
TITLE = {DisCaaS: Micro Behavior Analysis on Discussion by Camera as a Sensor},
JOURNAL = {Sensors},
VOLUME = {21},
YEAR = {2021},
NUMBER = {17},
ARTICLE-NUMBER = {5719},
URL = {https://www.mdpi.com/1424-8220/21/17/5719},
PubMedID = {34502609},
ISSN = {1424-8220},
DOI = {10.3390/s21175719}
}

@article{kahneman1966pupil,
  title     = {Pupil diameter and load on memory},
  author    = {Kahneman, Daniel and Beatty, Jackson},
  journal   = {Science},
  volume    = {154},
  number    = {3756},
  pages     = {1583--1585},
  year      = {1966},
  publisher = {American Association for the Advancement of Science}
}

@article{kucewicz2018pupil,
  title     = {Pupil size reflects successful encoding and recall of memory in humans},
  author    = {Kucewicz, Michal T and Dolezal, Jaromir and Kremen, Vaclav and Berry, Brent M and Miller, Laura R and Magee, Abigail L and Fabian, Vratislav and Worrell, Gregory A},
  journal   = {Scientific reports},
  volume    = {8},
  number    = {1},
  pages     = {4949},
  year      = {2018},
  publisher = {Nature Publishing Group UK London}
}

@inproceedings{kosch2018look,
author = {Kosch, Thomas and Hassib, Mariam and Buschek, Daniel and Schmidt, Albrecht},
title = {Look into my Eyes: Using Pupil Dilation to Estimate Mental Workload for Task Complexity Adaptation},
year = {2018},
isbn = {9781450356213},
publisher = {Association for Computing Machinery},
address = {New York, NY, USA},
url = {https://doi.org/10.1145/3170427.3188643},
doi = {10.1145/3170427.3188643},
booktitle = {Extended Abstracts of the 2018 CHI Conference on Human Factors in Computing Systems},
pages = {1–6},
numpages = {6},
location = {Montreal QC, Canada},
series = {CHI EA '18}
}

@inproceedings{kastrati2021using,
  title={Using deep learning to classify saccade direction from brain activity},
  author={Kastrati, Ard and Plomecka, Martyna Beata and Wattenhofer, Roger and Langer, Nicolas},
  booktitle={ACM Symposium on Eye Tracking Research and Applications},
  pages={1--6},
  year={2021}
}

@article{joshi2020pupil,
  title={Pupil size as a window on neural substrates of cognition},
  author={Joshi, Siddhartha and Gold, Joshua I},
  journal={Trends in cognitive sciences},
  volume={24},
  number={6},
  pages={466--480},
  year={2020},
  publisher={Elsevier}
}

@article{nobukawa2021pupillometric,
  title={Pupillometric complexity and symmetricity follow inverted-U curves against baseline diameter due to crossed locus coeruleus projections to the Edinger-Westphal nucleus},
  author={Nobukawa, Sou and Shirama, Aya and Takahashi, Tetsuya and Takeda, Toshinobu and Ohta, Haruhisa and Kikuchi, Mitsuru and Iwanami, Akira and Kato, Nobumasa and Toda, Shigenobu},
  journal={Frontiers in Physiology},
  volume={12},
  pages={614479},
  year={2021},
  publisher={Frontiers Media SA}
}

@inproceedings{grimmer2021cognitive,
author = {Grimmer, Janine and Simon, Laura and Ehlers, Jan},
title = {The Cognitive Eye: Indexing Oculomotor Functions for Mental Workload Assessment in Cognition-Aware Systems},
year = {2021},
isbn = {9781450380959},
publisher = {Association for Computing Machinery},
address = {New York, NY, USA},
url = {https://doi.org/10.1145/3411763.3451662},
doi = {10.1145/3411763.3451662},
booktitle = {Extended Abstracts of the 2021 CHI Conference on Human Factors in Computing Systems},
articleno = {428},
numpages = {6},
location = {Yokohama, Japan},
series = {CHI EA '21}
}

@inproceedings{hutt2021breaking,
author = {Hutt, Stephen and Krasich, Kristina and R. Brockmole, James and K. D'Mello, Sidney},
title = {Breaking out of the Lab: Mitigating Mind Wandering with Gaze-Based Attention-Aware Technology in Classrooms},
year = {2021},
isbn = {9781450380966},
publisher = {Association for Computing Machinery},
address = {New York, NY, USA},
url = {https://doi.org/10.1145/3411764.3445269},
doi = {10.1145/3411764.3445269},
booktitle = {Proceedings of the 2021 CHI Conference on Human Factors in Computing Systems},
articleno = {52},
numpages = {14},
location = {Yokohama, Japan},
series = {CHI '21}
}

@article{brishtel2020mind,
  title={Mind wandering in a multimodal reading setting: Behavior analysis \& automatic detection using eye-tracking and an eda sensor},
  author={Brishtel, Iuliia and Khan, Anam Ahmad and Schmidt, Thomas and Dingler, Tilman and Ishimaru, Shoya and Dengel, Andreas},
  journal={Sensors},
  volume={20},
  number={9},
  pages={2546},
  year={2020},
  publisher={MDPI}
}

@article{srivastava2018combining,
  title={Combining low and mid-level gaze features for desktop activity recognition},
  author={Srivastava, Namrata and Newn, Joshua and Velloso, Eduardo},
  journal={Proceedings of the ACM on Interactive, Mobile, Wearable and Ubiquitous Technologies},
  volume={2},
  number={4},
  pages={1--27},
  year={2018},
  publisher={ACM New York, NY, USA}
}

@article{walker1980lateralization,
  title={Lateralization of functions in the vertebrate brain: A review},
  author={Walker, SF},
  journal={British journal of psychology},
  volume={71},
  number={3},
  pages={329--367},
  year={1980},
  publisher={Wiley Online Library}
}

@article{neubauer2020evolution,
  title={Evolution of brain lateralization: A shared hominid pattern of endocranial asymmetry is much more variable in humans than in great apes},
  author={Neubauer, Simon and Gunz, Philipp and Scott, Nadia A and Hublin, Jean-Jacques and Mitteroecker, Philipp},
  journal={Science Advances},
  volume={6},
  number={7},
  pages={eaax9935},
  year={2020},
  publisher={American Association for the Advancement of Science}
}

@article{toga2003mapping,
  title={Mapping brain asymmetry},
  author={Toga, Arthur W and Thompson, Paul M},
  journal={Nature Reviews Neuroscience},
  volume={4},
  number={1},
  pages={37--48},
  year={2003},
  publisher={Nature Publishing Group UK London}
}

@article{semenza2006math,
  title={Is math lateralised on the same side as language? Right hemisphere aphasia and mathematical abilities},
  author={Semenza, Carlo and Delazer, Margarete and Bertella, Laura and Gran{\`a}, Alessia and Mori, Ileana and Conti, Fabio M and Pignatti, Riccardo and Bartha, Lisa and Domahs, Frank and Benke, Thomas and others},
  journal={Neuroscience Letters},
  volume={406},
  number={3},
  pages={285--288},
  year={2006},
  publisher={Elsevier}
}

@article{pinel2010beyond,
  title={Beyond hemispheric dominance: brain regions underlying the joint lateralization of language and arithmetic to the left hemisphere},
  author={Pinel, Philippe and Dehaene, Stanislas},
  journal={Journal of Cognitive Neuroscience},
  volume={22},
  number={1},
  pages={48--66},
  year={2010},
  publisher={MIT Press One Rogers Street, Cambridge, MA 02142-1209, USA journals-info~…}
}

@article{kirk1989hemispheric,
  title={Hemispheric contributions to drawing},
  author={Kirk, Andrew and Kertesz, Andrew},
  journal={Neuropsychologia},
  volume={27},
  number={6},
  pages={881--886},
  year={1989},
  publisher={Elsevier}
}

@book{gordon1983music,
  title={Music and the right hemisphere},
  author={Gordon, H},
  volume={3},
  year={1983},
  publisher={Academic Press London}
}

@article{peretz2005brain,
  title={Brain organization for music processing},
  author={Peretz, Isabelle and Zatorre, Robert J},
  journal={Annu. Rev. Psychol.},
  volume={56},
  number={1},
  pages={89--114},
  year={2005},
  publisher={Annual Reviews}
}

@inproceedings{cho2021rethinking,
author = {Cho, Youngjun},
title = {Rethinking Eye-blink: Assessing Task Difficulty through Physiological Representation of Spontaneous Blinking},
year = {2021},
isbn = {9781450380966},
publisher = {Association for Computing Machinery},
address = {New York, NY, USA},
url = {https://doi.org/10.1145/3411764.3445577},
doi = {10.1145/3411764.3445577},
booktitle = {Proceedings of the 2021 CHI Conference on Human Factors in Computing Systems},
articleno = {721},
numpages = {12},
location = {Yokohama, Japan},
series = {CHI '21}
}

@inproceedings{duchowski2018task,
author = {Duchowski, Andrew T. and Krejtz, Krzysztof and Krejtz, Izabela and Biele, Cezary and Niedzielska, Anna and Kiefer, Peter and Raubal, Martin and Giannopoulos, Ioannis},
title = {The Index of Pupillary Activity: Measuring Cognitive Load vis-\`{a}-vis Task Difficulty with Pupil Oscillation},
year = {2018},
isbn = {9781450356206},
publisher = {Association for Computing Machinery},
address = {New York, NY, USA},
url = {https://doi.org/10.1145/3173574.3173856},
doi = {10.1145/3173574.3173856},
booktitle = {Proceedings of the 2018 CHI Conference on Human Factors in Computing Systems},
pages = {1–13},
numpages = {13},
location = {Montreal QC, Canada},
series = {CHI '18}
}

@inproceedings{bacchin2023gaze,
author = {Bacchin, Davide and Gehrer, Nina A. and Krejtz, Krzysztof and Duchowski, Andrew T. and Gamberini, Luciano},
title = {Gaze-based Metrics of Cognitive Load in a Conjunctive Visual Memory Task},
year = {2023},
isbn = {9781450394222},
publisher = {Association for Computing Machinery},
address = {New York, NY, USA},
url = {https://doi.org/10.1145/3544549.3585650},
doi = {10.1145/3544549.3585650},
booktitle = {Extended Abstracts of the 2023 CHI Conference on Human Factors in Computing Systems},
articleno = {148},
numpages = {8},
location = {Hamburg, Germany},
series = {CHI EA '23}
}

@inproceedings{tag2019continuous,
author = {Tag, Benjamin and Vargo, Andrew W. and Gupta, Aman and Chernyshov, George and Kunze, Kai and Dingler, Tilman},
title = {Continuous Alertness Assessments: Using EOG Glasses to Unobtrusively Monitor Fatigue Levels In-The-Wild},
year = {2019},
isbn = {9781450359702},
publisher = {Association for Computing Machinery},
address = {New York, NY, USA},
url = {https://doi.org/10.1145/3290605.3300694},
doi = {10.1145/3290605.3300694},
booktitle = {Proceedings of the 2019 CHI Conference on Human Factors in Computing Systems},
pages = {1–12},
numpages = {12},
location = {Glasgow, Scotland Uk},
series = {CHI '19}
}

@inproceedings{abdulin2015user,
author = {Abdulin, Evgeniy and Komogortsev, Oleg},
title = {User Eye Fatigue Detection via Eye Movement Behavior},
year = {2015},
isbn = {9781450331463},
publisher = {Association for Computing Machinery},
address = {New York, NY, USA},
url = {https://doi.org/10.1145/2702613.2732812},
doi = {10.1145/2702613.2732812},
booktitle = {Proceedings of the 33rd Annual ACM Conference Extended Abstracts on Human Factors in Computing Systems},
pages = {1265–1270},
numpages = {6},
location = {Seoul, Republic of Korea},
series = {CHI EA '15}
}

@article{mann2002suspects,
  title={Suspects, lies, and videotape: An analysis of authentic high-stake liars},
  author={Mann, Samantha and Vrij, Aldert and Bull, Ray},
  journal={Law and human behavior},
  volume={26},
  pages={365--376},
  year={2002},
  publisher={Springer}
}

@article{leal2008blinking,
  title={Blinking during and after lying},
  author={Leal, Sharon and Vrij, Aldert},
  journal={Journal of Nonverbal Behavior},
  volume={32},
  pages={187--194},
  year={2008},
  publisher={Springer}
}

@inproceedings{grootjen2024uncovering,
author = {Grootjen, Jesse W. and Weing\"{a}rtner, Henrike and Mayer, Sven},
title = {Uncovering and Addressing Blink-Related Challenges in Using Eye Tracking for Interactive Systems},
year = {2024},
isbn = {9798400703300},
publisher = {Association for Computing Machinery},
address = {New York, NY, USA},
url = {https://doi.org/10.1145/3613904.3642086},
doi = {10.1145/3613904.3642086},
booktitle = {Proceedings of the CHI Conference on Human Factors in Computing Systems},
articleno = {322},
numpages = {23},
location = {Honolulu, HI, USA},
series = {CHI '24}
}

@article{rogers2021brain,
  title={Brain lateralization and cognitive capacity},
  author={Rogers, Lesley J},
  journal={Animals},
  volume={11},
  number={7},
  pages={1996},
  year={2021},
  publisher={MDPI}
}

@inproceedings{xu2011pupillary,
  title={Pupillary response based cognitive workload measurement under luminance changes},
  author={Xu, Jie and Wang, Yang and Chen, Fang and Choi, Eric},
  booktitle={IFIP Conference on Human-Computer Interaction},
  pages={178--185},
  year={2011},
  organization={Springer}
}

@article{daguet2019baseline,
  title={Baseline pupil diameter is not a reliable biomarker of subjective sleepiness},
  author={Daguet, In{\`e}s and Bouhassira, Didier and Gronfier, Claude},
  journal={Frontiers in Neurology},
  volume={10},
  pages={108},
  year={2019},
  publisher={Frontiers Media SA}
}

@article{mathot2018safe,
  title={Safe and sensible preprocessing and baseline correction of pupil-size data},
  author={Math{\^o}t, Sebastiaan and Fabius, Jasper and Van Heusden, Elle and Van der Stigchel, Stefan},
  journal={Behavior research methods},
  volume={50},
  pages={94--106},
  year={2018},
  publisher={Springer}
}

@article{yau2021evidence,
  title={Evidence and urgency related EEG signals during dynamic decision-making in humans},
  author={Yau, Yvonne and Hinault, Thomas and Taylor, Madeline and Cisek, Paul and Fellows, Lesley K and Dagher, Alain},
  journal={Journal of Neuroscience},
  volume={41},
  number={26},
  pages={5711--5722},
  year={2021},
  publisher={Soc Neuroscience}
}

@article{lim2022eye,
  title={Eye-tracking feature extraction for biometric machine learning},
  author={Lim, Jia Zheng and Mountstephens, James and Teo, Jason},
  journal={Frontiers in neurorobotics},
  volume={15},
  pages={796895},
  year={2022},
  publisher={Frontiers Media SA}
}

@article{liversedge2000saccadic,
  title={Saccadic eye movements and cognition},
  author={Liversedge, Simon P and Findlay, John M},
  journal={Trends in cognitive sciences},
  volume={4},
  number={1},
  pages={6--14},
  year={2000},
  publisher={Elsevier}
}

@inproceedings{bulling2009eye,
  title={Eye movement analysis for activity recognition},
  author={Bulling, Andreas and Ward, Jamie A and Gellersen, Hans and Tr{\"o}ster, Gerhard},
  booktitle={Proceedings of the 11th international conference on Ubiquitous computing},
  pages={41--50},
  year={2009}
}

@article{vortmann2021combining,
  title={Combining implicit and explicit feature extraction for eye tracking: attention classification using a heterogeneous input},
  author={Vortmann, Lisa-Marie and Putze, Felix},
  journal={Sensors},
  volume={21},
  number={24},
  pages={8205},
  year={2021},
  publisher={MDPI}
}

@misc{PupilLabsCore,
  author       = {PupilLabs},
  title        = {Specifications for the Pupil Labs Core Eye Tracker},
  howpublished = {\url{https://pupil-labs.com/products/core}},
  address      = {Berlin, Germany},
  year         = {2024}
}

@misc{PupilLabsInvisible,
  author       = {PupilLabs},
  title        = {Specifications for the Pupil Labs Invisible Eye Tracker},
  howpublished = {\url{https://pupil-labs.com/products/invisible}},
  address      = {Berlin, Germany},
  year         = {2024}
}

@article{aminihajibashi2019individual,
  title={Individual differences in resting-state pupil size: Evidence for association between working memory capacity and pupil size variability},
  author={Aminihajibashi, Samira and Hagen, Thomas and Foldal, Maja Dyhre and Laeng, Bruno and Espeseth, Thomas},
  journal={International Journal of Psychophysiology},
  volume={140},
  pages={1--7},
  year={2019},
  publisher={Elsevier}
}

@article{kret2019preprocessing,
  title={Preprocessing pupil size data: Guidelines and code},
  author={Kret, Mariska E and Sjak-Shie, Elio E},
  journal={Behavior research methods},
  volume={51},
  pages={1336--1342},
  year={2019},
  publisher={Springer}
}

@article{tanaka2022sliding,
  title={Sliding-window normalization to improve the performance of machine-learning models for real-time motion prediction using electromyography},
  author={Tanaka, Taichi and Nambu, Isao and Maruyama, Yoshiko and Wada, Yasuhiro},
  journal={Sensors},
  volume={22},
  number={13},
  pages={5005},
  year={2022},
  publisher={MDPI}
}

@inproceedings{ishimaru2014blink,
  title={In the blink of an eye: combining head motion and eye blink frequency for activity recognition with google glass},
  author={Ishimaru, Shoya and Kunze, Kai and Kise, Koichi and Weppner, Jens and Dengel, Andreas and Lukowicz, Paul and Bulling, Andreas},
  booktitle={Proceedings of the 5th augmented human international conference},
  pages={1--4},
  year={2014}
}

@incollection{rogers2024lateralization,
  title={Lateralization of Brain Function},
  author={Rogers, Lesley J},
  booktitle={Oxford Research Encyclopedia of Psychology},
  year={2024}
}

@book{ocklenburg2024lateralized,
  title={The lateralized brain: The neuroscience and evolution of hemispheric asymmetries},
  author={Ocklenburg, Sebastian and G{\"u}nt{\"u}rk{\"u}n, Onur},
  year={2024},
  publisher={Elsevier}
}

@article{dapretto1999form,
  title={Form and content: dissociating syntax and semantics in sentence comprehension},
  author={Dapretto, Mirella and Bookheimer, Susan Y},
  journal={Neuron},
  volume={24},
  number={2},
  pages={427--432},
  year={1999},
  publisher={Elsevier}
}

@misc{binder2000new,
  title={The new neuroanatomy of speech perception},
  author={Binder, Jeffrey},
  journal={Brain},
  volume={123},
  number={12},
  pages={2371--2372},
  year={2000},
  publisher={Oxford University Press}
}

@article{kong2018mapping,
  title={Mapping cortical brain asymmetry in 17,141 healthy individuals worldwide via the ENIGMA Consortium},
  author={Kong, Xiang-Zhen and Mathias, Samuel R and Guadalupe, Tulio and ENIGMA Laterality Working Group and Glahn, David C and Franke, Barbara and Crivello, Fabrice and Tzourio-Mazoyer, Nathalie and Fisher, Simon E and Thompson, Paul M and others},
  journal={Proceedings of the National Academy of Sciences},
  volume={115},
  number={22},
  pages={E5154--E5163},
  year={2018},
  publisher={National Acad Sciences}
}

@article{kong2022mapping,
  title={Mapping brain asymmetry in health and disease through the ENIGMA consortium},
  author={Kong, Xiang-Zhen and Postema, Merel C and Guadalupe, Tulio and de Kovel, Carolien and Boedhoe, Premika SW and Hoogman, Martine and Mathias, Samuel R and Van Rooij, Daan and Schijven, Dick and Glahn, David C and others},
  journal={Human brain mapping},
  volume={43},
  number={1},
  pages={167--181},
  year={2022},
  publisher={Wiley Online Library}
}

@article{funnell2007calculating,
  title={The calculating hemispheres: Studies of a split-brain patient},
  author={Funnell, Margaret G and Colvin, Mary K and Gazzaniga, Michael S},
  journal={Neuropsychologia},
  volume={45},
  number={10},
  pages={2378--2386},
  year={2007},
  publisher={Elsevier}
}

@article{nikolaenko2003artistic,
  title={Artistic thinking and cerebral asymmetry},
  author={Nikolaenko, Nikolai},
  journal={Acta Neuropsychologica},
  volume={1},
  number={2},
  year={2003},
  publisher={Index Copernicus}
}

@inproceedings{putze2020platform,
  title={Platform for studying self-repairing auto-corrections in mobile text entry based on brain activity, gaze, and context},
  author={Putze, Felix and Ihrig, Tilman and Schultz, Tanja and Stuerzlinger, Wolfgang},
  booktitle={Proceedings of the 2020 CHI Conference on Human Factors in Computing Systems},
  pages={1--13},
  year={2020}
}

@inproceedings{reddy2024towards,
  title={Towards an Eye-Brain-Computer Interface: Combining Gaze with the Stimulus-Preceding Negativity for Target Selections in XR},
  author={Reddy, GS Rajshekar and Proulx, Michael J and Hirshfield, Leanne and Ries, Anthony},
  booktitle={Proceedings of the CHI Conference on Human Factors in Computing Systems},
  pages={1--17},
  year={2024}
}

@inproceedings{berkovsky2019detecting,
author = {Berkovsky, Shlomo and Taib, Ronnie and Koprinska, Irena and Wang, Eileen and Zeng, Yucheng and Li, Jingjie and Kleitman, Sabina},
title = {Detecting Personality Traits Using Eye-Tracking Data},
year = {2019},
isbn = {9781450359702},
publisher = {Association for Computing Machinery},
address = {New York, NY, USA},
url = {https://doi.org/10.1145/3290605.3300451},
doi = {10.1145/3290605.3300451},
booktitle = {Proceedings of the 2019 CHI Conference on Human Factors in Computing Systems},
pages = {1–12},
numpages = {12},
location = {Glasgow, Scotland Uk},
series = {CHI '19}
}

@InProceedings{Hisadome_2024_WACV,
    author    = {Hisadome, Yoichiro and Wu, Tianyi and Qin, Jiawei and Sugano, Yusuke},
    title     = {Rotation-Constrained Cross-View Feature Fusion for Multi-View Appearance-Based Gaze Estimation},
    booktitle = {Proceedings of the IEEE/CVF Winter Conference on Applications of Computer Vision (WACV)},
    month     = {January},
    year      = {2024},
    pages     = {5985-5994}
}

@inproceedings{zhang2019evaluation,
author = {Zhang, Xucong and Sugano, Yusuke and Bulling, Andreas},
title = {Evaluation of Appearance-Based Methods and Implications for Gaze-Based Applications},
year = {2019},
isbn = {9781450359702},
publisher = {Association for Computing Machinery},
address = {New York, NY, USA},
url = {https://doi.org/10.1145/3290605.3300646},
doi = {10.1145/3290605.3300646},
booktitle = {Proceedings of the 2019 CHI Conference on Human Factors in Computing Systems},
pages = {1–13},
numpages = {13},
location = {Glasgow, Scotland Uk},
series = {CHI '19}
}

@inproceedings{housholder2021evaluating,
  title={Evaluating accuracy of the Tobii eye tracker 5},
  author={Housholder, Andrew and Reaban, Jonathan and Peregrino, Aira and Votta, Georgia and Mohd, Tauheed Khan},
  booktitle={International Conference on Intelligent Human Computer Interaction},
  pages={379--390},
  year={2021},
  organization={Springer}
}

@article{krejtz2018eye,
  title={Eye tracking cognitive load using pupil diameter and microsaccades with fixed gaze},
  author={Krejtz, Krzysztof and Duchowski, Andrew T and Niedzielska, Anna and Biele, Cezary and Krejtz, Izabela},
  journal={PloS one},
  volume={13},
  number={9},
  pages={e0203629},
  year={2018},
  publisher={Public Library of Science San Francisco, CA USA}
}

@article{oliva2018pupil,
  title={Pupil dilation reflects the time course of emotion recognition in human vocalizations},
  author={Oliva, Manuel and Anikin, Andrey},
  journal={Scientific reports},
  volume={8},
  number={1},
  pages={4871},
  year={2018},
  publisher={Nature Publishing Group UK London}
}

@inproceedings{ning2023smartphone,
author = {Ning, Emma and Cladek, Andrea T. and Ross, Mindy K. and Kabir, Sarah and Barve, Amruta and Kennelly, Ellyn and Hussain, Faraz and Duffecy, Jennifer and Langenecker, Scott L. and Nguyen, Theresa and Tulabandhula, Theja and Zulueta, John and Ajilore, Olusola A. and Demos, Alexander P. and Leow, Alex},
title = {Smartphone-derived Virtual Keyboard Dynamics Coupled with Accelerometer Data as a Window into Understanding Brain Health: Smartphone Keyboard and Accelerometer as Window into Brain Health},
year = {2023},
isbn = {9781450394215},
publisher = {Association for Computing Machinery},
address = {New York, NY, USA},
url = {https://doi.org/10.1145/3544548.3580906},
doi = {10.1145/3544548.3580906},
booktitle = {Proceedings of the 2023 CHI Conference on Human Factors in Computing Systems},
articleno = {326},
numpages = {15},
location = {Hamburg, Germany},
series = {CHI '23}
}

@inproceedings{gjoreski2023ocosense,
author = {Gjoreski, Hristijan and Mavridou, Ifigeneia and Archer, James Archer William and Cleal, Andrew and Stankoski, Simon and Kiprijanovska, Ivana and Fatoorechi, Mohsen and Walas, Piotr and Broulidakis, John and Gjoreski, Martin and Nduka, Charles},
title = {OCOsense Glasses – Monitoring Facial Gestures and Expressions for Augmented Human-Computer Interaction: OCOsense Glasses for Monitoring Facial Gestures and Expressions},
year = {2023},
isbn = {9781450394222},
publisher = {Association for Computing Machinery},
address = {New York, NY, USA},
url = {https://doi.org/10.1145/3544549.3583918},
doi = {10.1145/3544549.3583918},
booktitle = {Extended Abstracts of the 2023 CHI Conference on Human Factors in Computing Systems},
articleno = {465},
numpages = {4},
location = {Hamburg, Germany},
series = {CHI EA '23}
}

@inproceedings{yan2022emoglass,
author = {Yan, Zihan and Wu, Yufei and Zhang, Yang and Chen, Xiang 'Anthony'},
title = {EmoGlass: an End-to-End AI-Enabled Wearable Platform for Enhancing Self-Awareness of Emotional Health},
year = {2022},
isbn = {9781450391573},
publisher = {Association for Computing Machinery},
address = {New York, NY, USA},
url = {https://doi.org/10.1145/3491102.3501925},
doi = {10.1145/3491102.3501925},
booktitle = {Proceedings of the 2022 CHI Conference on Human Factors in Computing Systems},
articleno = {13},
numpages = {19},
location = {New Orleans, LA, USA},
series = {CHI '22}
}

@article{watanabe2023engauge,
  title={Engauge: Engagement gauge of meeting participants estimated by facial expression and deep neural network},
  author={Watanabe, Ko and Sathyanarayana, Tanuja and Dengel, Andreas and Ishimaru, Shoya},
  journal={IEEE Access},
  volume={11},
  pages={52886--52898},
  year={2023},
  publisher={IEEE}
}

@article{rozado2015combining,
  title={Combining EEG with pupillometry to improve cognitive workload detection},
  author={Rozado, David and Dunser, Andreas},
  journal={Computer},
  volume={48},
  number={10},
  pages={18--25},
  year={2015},
  publisher={IEEE}
}

@article{morita2020right,
  title={Right-hemispheric dominance in self-body recognition is altered in left-handed individuals},
  author={Morita, Tomoyo and Asada, Minoru and Naito, Eiichi},
  journal={Neuroscience},
  volume={425},
  pages={68--89},
  year={2020},
  publisher={Elsevier}
}

@article{corballis2014left,
  title={Left brain, right brain: facts and fantasies},
  author={Corballis, Michael C},
  journal={PLoS biology},
  volume={12},
  number={1},
  pages={e1001767},
  year={2014},
  publisher={Public Library of Science San Francisco, USA}
}

\appendix

\end{document}